# High stability white light generation in water at multi-kilohertz repetition rate

KILIAN RICHARD KELLER,[1,+] RICARDO ROJAS-AEDO,[1,+] ALINE VANDERHAEGEN,[1] MARKUS LUDWIG,[2] AND DANIELE BRIDA [1,*]

[1]*Université du Luxembourg, 162a, avenue de la Faïencerie, L-1511 Luxembourg, Luxembourg*
[2] *Deutsches Elektronen-Synchrotron DESY, Notkestr. 85, 22607 Hamburg, Germany*
*[daniele.brida@uni.lu](mailto:daniele.brida@uni.lu)*
[+] *K.R.K. and R.R.-A. contributed equally to this work.*

**Abstract:** Efficient supercontinuum (SC) generation featuring high spectral intensity across a large bandwidth requires high peak powers of several megawatt from pulsed lasers. Under these conditions and at multi-kilohertz (kHz) repetition rates, the SC generated in most materials is unstable due to thermal effects. In this work, we leverage the superior dispersion properties of water to maximize the spectral width of the SC, while avoiding stability issues due to thermal loading by means of a constant laminar flow of the liquid. This flow is controlled by a differential pressure scheme that allows to precisely adjust the fluid velocity to an optimum value for maximum stability of the SC. This approach is successfully implemented for repetition rates of 50 kHz and 100 kHz and two different pump wavelengths in the visible (VIS) and near infrared (NIR) spectral region with stability of the SC signal only limited by the driving pulses. The resulting water SC spans more than one octave covering the VIS to NIR range. Compared to established materials, such as yttrium aluminum garnet (YAG) and sapphire, the spectral bandwidth is increased by 60 % and 40 % respectively. Our scheme has the potential to be implemented with other liquids such as bromine or carbon disulfide ($CS_2$), which promise even wider broadening and operation up to the mid-infrared.

## 1. Introduction

The development of laser technology that provides increasingly shorter optical pulses in different spectral ranges is essential for ultrafast spectroscopy research and is widely based on the generation of SC in dielectric media [1,2]. The nonlinear nature of SC generation requires high peak pulse intensities, typically on the order of hundreds of GW cm$^{-2}$, which often induce thermal effects in the material. In general, these effects can be problematic if the thermal relaxation time of the medium exceeds the time separation between consecutive optical pulses, which impacts the short and long term stability of the SC. Short term stability effects may be caused by thermal lensing or thermal blooming [3], which changes the geometry of the plasma filament and thus the spectrum of the SC and the generation efficiency. While in many cases this effect can be neglected due to the low dependence of the refractive index on temperature, there are other long-term effects that cannot be neglected, such as optical degradation and irreversible damage in solid materials [4], which becomes especially problematic at multi-kHz repetition rates with pump pulses with a duration of few hundreds of femtoseconds [1]. For ytterbium (Yb)-based laser amplifiers operating in this regime, YAG crystals provide the best stability [5,6], and under extreme conditions YAG is even considered the only reliable material [1]. Efforts to overcome the degradation at multi-kHz repetition rates in alternative crystals which allow for more efficient spectral broadening in the ultraviolet (UV)-

VIS spectral region, such as the rather delicate $CaF_2$ [7], are therefore the subject of current research. In that regard, liquids are an attractive alternative to solids since they do not suffer from permanent optical damage. Despite this strong advantage there is a clear gap in scientific efforts with respect to SC generation in liquids at multi-kHz repetition rates, which we attribute to the formation of bubbles or shock waves in this regime that drastically reduce the stability of a SC signal. The lack of studies investigating these instabilities at multi-kHz repetition rates to find solutions may be related to an established view that solids have replaced liquids as a source of bulk SC generation due to their alleged higher efficiency and stability [8]. The former assumption is challenged by the fact that water, despite its relatively low nonlinear refractive index compared to other media, produces one of the broadest anti-Stokes spectra when pumped at 800 nm [9], which is attributed to its high band gap (7.5 eV [10]) and low group velocity dispersion (58 $fs^2$/mm at 515 nm [11]). Furthermore, recent studies demonstrate SC generation in water spanning over two octaves from the UV to the NIR spectral region [12,13] at 1 kHz repetition rate. The stability problems arising due to thermal loading induced by pump pulses at repetition rates above 1 kHz, are attributed to the low thermal conductivity of water (0.6 W/mK) [14], which is about two orders of magnitude below other materials commonly used in SC generation such as sapphire (35 W/(mK)) [15] and YAG (12 W/(mK)) [16]. Figure 1 a) shows that stable SC can be generated in water inside a liquid cell when pumped at 1 kHz by the second harmonic (SH) (515nm) of a commercial Yb:KGW laser amplifier. However, as the repetition rate increases to tens of kilohertz, the signal becomes increasingly unstable, with the occurrence of significant beam breakups (inset – red arrow). Between these breakups, the SC signal is stable for the duration of seconds to tens of seconds for 10 kHz and from hundreds of milliseconds (zoom of inset) to seconds at 50 kHz. Since higher repetition rates imply shorter timescales for thermal relaxation, the signal breakup is clearly induced by an accumulation of heat and the consequent formation of bubbles. On the other hand, if a precisely controlled flow of water is induced, the signal at 50 kHz can become even more stable than the signal at 1 kHz without flow, as shown in Fig. 1 b).

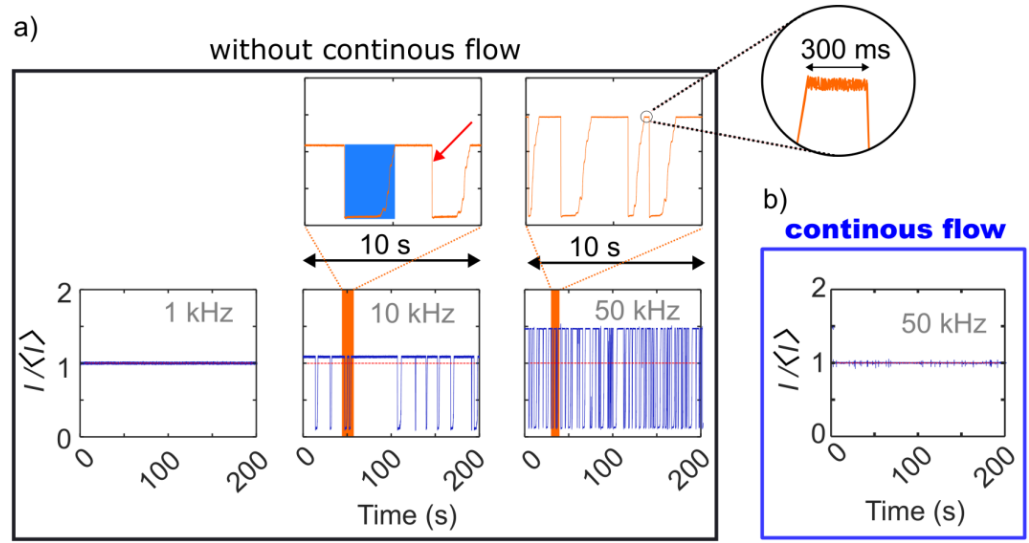

Fig. 1. Pulse-to-pulse fluctuation of SC signal generated in water a) without continuous flow at 1 kHz, 10 kHz and 50 kHz with a peak pump intensity of 170 GW $cm^{-2}$, 290 GW $cm^{-2}$ and 310 GW $cm^{-2}$ respectively and b) with a controlled continuous flow at 50 kHz with a peak pump intensity of 690 GW $cm^{-2}$. The SC is pumped by the SH at 515 nm and focused by a f = 50 mm lens. The inset of the measurements performed at 10 kHz and 50 kHz highlights the signal recovery (blue area) and the sharp decays (red arrow).

This approach, although effective in eliminating strong thermal instabilities, requires a particularly precise optimization of the fluid velocity in order to reduce thermal effects while avoiding excessive beam pointing instabilities due to turbulent flow. The first part of the paper presents the solution developed to precisely control differential pressure and thus the fluid velocity of water interacting with the optical pulse, which allows to generate stable octave-spanning SC signals in water at multi-kHz repetition rates. The following part includes a detailed comparison of spectra and stability of the SC generated in water, YAG and sapphire when pumped by the SH and by the fundamental wavelength (FW) of an Yb-based femtosecond system. In the last part, the stability of SC generation in water is measured as a function of fluid velocity for 50 kHz and 100 kHz.

## 2. Experimental Setup

In our experiment we employ a Yb:KGW laser amplifier (PHAROS from Light Conversion) that provides pulses at a repetition rate tunable from 1 kHz to 100 kHz with a pulse duration of 230 fs at a central wavelength of the FW at 1030 nm. The second harmonic (SH) of these pulses is generated in a β-barium borate (BBO) crystal with a pulse duration of 160 fs, where the process is driven in saturation in terms of conversion efficiency for optimum stability. The SH and FW pulses are then used to drive SC generation in distilled water inside a liquid cell. The experimental setup is shown in Fig. 2. The cell consists of a cuvette sealed with 100 μm thick sapphire windows, which are separated by 5.6 mm. Sapphire is chosen in particular for its high optical damage threshold [17]. Inside the cell, the water is indirectly driven via a pressure difference between an open canister (C1) and a pressurized closed canister (C2) connected by a water pump (P). The pressure difference is set with a bypass valve (BV) attached to a tube, that connects the two canisters. The flow rate inside the cell is fine-tuned by another valve (V) attached to the tube connecting the cell and the open canister. The combination of these two valves is crucial to fine tune the differential pressure between the canisters and thus precisely set the fluid velocity inside the cell. Furthermore, the rather large volume (one liter) of the canister C2 helps to reduce mechanical pressure fluctuations of the pump, which would destabilize the pressure that drives the water through the cell. Other sources of instability such as particles suspended in the water and persistent bubbles are eliminated with a water filter (WF) and a bubble trap [18], respectively. In the experiments, the pulses centered at 515 nm with a beam waist of 2 mm pass through an iris (I) set 11 mm before a f = 50 mm lens (LA1131-A, Thorlabs). The diffraction due to the iris enables redistribution of the beam intensity profile as shown in Fig. 2 b), which optimizes the signal stability and also the bandwidth of the SC [19,20]. Optimal conditions are achieved under loose focusing with an iris aperture diameter of about 0.6 mm, which corresponds to an numerical aperture (NA) of 0.01. With the measured full-width-at-half-maximum (FWHM) spot size at the geometric focus of 31 μm in free space this results in a peak pulse intensity of 690 GW cm$^{-2}$. The pump power is adjusted by a combination of half-wave plate and polarizing beam splitter cube. We set the geometrical focus slightly before the input window of the cell, which in combination with a low NA leads to a more efficient SC generation [21] and an enhanced red-shifted broadening [22]. After SC generation, the Stokes and the anti-Stokes shoulder of the SC are separated from the SH pump by a long-pass with cut-on wavelength at 530 nm and a short-pass with cut-off wavelength at 500 nm, respectively. The spectra are analyzed with an intensity calibrated ultraviolet extended spectrometer (OceanOptics USB2000+), while a silicon photodiode (S5972, Hamamatsu) is used for measuring the pulse-to-pulse stability.
The SC generation in water is compared to a SC generated in a YAG (111-cut) and a sapphire (0001-cut) crystal of 6-mm thickness. To calibrate the spectral intensity and compare relative measurements, the average power of the spectra is measured after a band pass with a FWHM of 10 nm. The pulse-to-pulse stability is then characterized by means of a boxcar detection scheme (UHFLI, Zurich Instruments) at the full repetition rate of the driving pulses. All SC signals have been optimized for maximum spectral broadening while maintaining a root mean square (rms) pulse-to-pulse stability close to limit set by the pump laser. The focal spot size for both the crystals and the water cell are identical, resulting in peak pulse intensities of 344 GW cm$^{-2}$ and 826 GW cm$^{-2}$ for YAG and sapphire, respectively. For the characterization of the SC pumped at 1030 nm, the FW of the laser is focused with a f = 100 mm lens. The focal length is doubled with respect to SH pumping to ensure similar focusing conditions. The geometrical focus is now set inside the water to reduce the energy threshold for SC generation necessary for self-focusing at longer wavelengths [1]. Due to the absorption of water in the infrared, only the anti-Stokes shoulder is analyzed by using a short-pass filter with cut-off wavelength at 900 nm. The optimized peak pump intensities are 1640 GW cm$^{-2}$, 420 GW cm$^{-2}$ and 820 GW cm$^{-2}$ for water, YAG and sapphire, respectively.

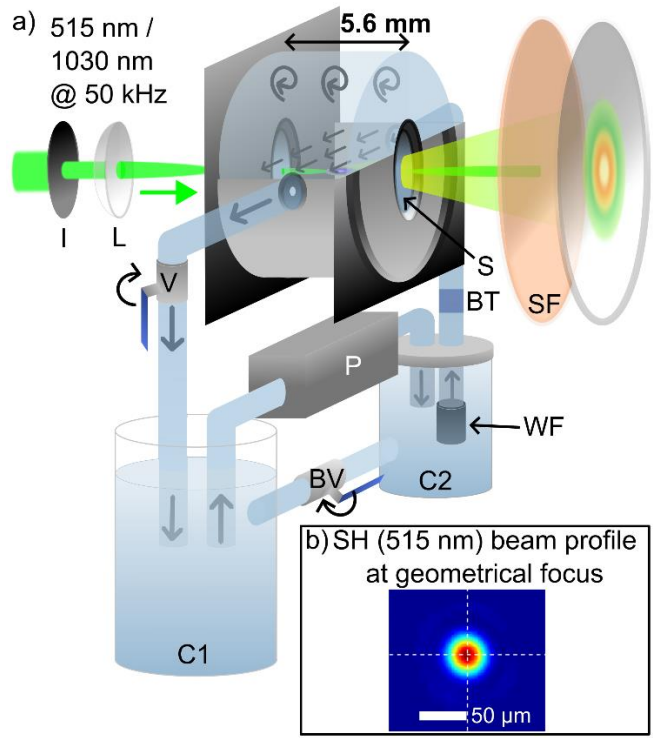

Fig. 2. a) Setup for the generation of stable SC in water at multi-kHz repetition rates. The SC is pumped alternately by the SH (515 nm) and the FW (1030 nm) of a Yb:KGW laser amplifier. The light of the SH and FW passes through an iris (I) and is focused into the water inside a liquid cell with a lens (L) of focal length f = 50 mm and f = 100 mm, for the two cases. The two sapphire windows (S, C-plane cut, thickness 100 µm) of the flow cell are separated by a distance of 5.6 mm. An in- and outlet with a diameter of approximately 1.5 mm allows to pump liquids through the cell, which is connected via tubes to two canisters filled with distilled water. Water is transferred with a diaphragm pump (P) from an open canister C1 to a closed canister C2, where pressure can build up. A first bypass valve (BV) that connects the two canisters and a second valve (V) attached to the tube in between the cell and the open canister allow to fine adjust the differential pressure between the canisters and thus the fluid velocity inside the cell. The water passes a particle filter (WF) and a bubble trap (BT) [18] before entering the cell. A spectral filter (SF) blocks the pump wavelength from the generated SC after the cell. b) Beam profile of the SH pump at the geometrical focus.

## 3. Results and Discussion

In general maximum spectral broadening and optimum stability impose constraints on the peak power of the pump pulse with respect to the physical characteristics and dimensions of the nonlinear material. While initially increasing the pump power results in spectral broadening of the SC, optimum stability is achieved when the broadening of the anti-Stokes signal is saturated. Further increase of the pump power beyond this saturation results in additional broadening on the Stokes side [8] up to the point when additional effects emerge that influence pulse propagation and can reduce transmission and stability. One example is that under loose focusing conditions, if the power is increased to a few times the critical power needed for a self-focusing, the pulse undergoes catastrophic collapse ($P_{crit}$) and a recurrent spatial refocusing of the filament takes place. The occurrence of filament refocusing depends furthermore on the length of the medium and is characterized by the appearance of interference fringes in the spectrum. The peak pump power that optimizes spectral width while preserving high stability and temporal signal structure for a fixed length of the nonlinear medium is reached at the onset of refocusing. Consequently, the optimum intensity for the pump can be found experimentally with the appearance of small fringes in the SC spectrum. For water in the flow cell these fringes appear close to the driving wavelength at a peak pump power of 7.5 MW (Fig. 3 (a.1) and b.1). This value is almost 4.5 times the value of $P_{crit} \approx 1.7$ MW for water. $P_{crit}$ is calculated using the linear and nonlinear refractive indices of $n = 1.34$ [11] and $n_2 = 1.7 \cdot 10^{-16}$ cm$^2$/W [23] respectively, where $n_2$ is determined at 407 nm and thus serves as an approximation of the value at the wavelength of the SH. Similarly, the optimum peak power of the FW to generate a SC in water is 20 MW, which corresponds to about 3 times the critical power of $P_{crit} \approx 6.3$ MW calculated with a nonlinear refractive index of $n_2 = 1.9 \cdot 10^{-16}$ cm$^2$/W [23], where $n_2$ is determined at 815 nm and thus serves as an approximation of the value to the wavelength of the FW. To compare the spectral broadening between water and the dielectric crystals we define the wavelengths $\lambda_{5\%,ST}$ and $\lambda_{5\%,AS}$ at which the spectral intensity drops to 5% with respect to the maximum signal for the stokes and anti-stokes side respectively. Similarly, to compare the spectral power for different materials we define the average power ($P_{avg}$) as the signal power after a bandpass in different spectral regions. For the white light generated with the SH (Fig. 3 a.1) and b.1)) we determined $P_{avg}$ after a bandpass centered at 700 nm and 450 nm for the Stokes and anti-Stokes signal, respectively. For the white light generated with the FW (Fig. 3 c.1)) a bandpass centered at 671 nm is used to measure $P_{avg}$ of the anti-Stokes signal. The optimized spectra of YAG and sapphire (Fig. 3 a.1, b.1 and c.1) are in good agreement with results presented by A. L. Calendron et al. [24] for comparable focusing conditions and the same pump wavelengths. The $P_{avg}$ and $\lambda_{5\%}$ for each case are summarized in table 1. Among the crystals, sapphire provides a higher $P_{avg}$ on both the anti-Stokes and Stokes side of the SC. When pumped with the SH, $P_{avg}$ of the SC from water is 6 and 3 times higher than for sapphire on the anti-Stokes and Stokes sides respectively. Furthermore water provides an octave-spanning SC, and compared to YAG and sapphire, a bandwidth enhancement of 60 % and 40 % respectively. For the FW pump, the $P_{avg}$ at 671 nm is comparable for all three materials. The calibrated spectrum (Fig. 3 c.1)) shows that around $\lambda_{5\%}$ of the crystals at 500 nm the spectral density in water is more than twice that of sapphire, with an octave spanning anti-Stokes shoulder that extends more than 10% further than in the crystals.

The pulse-to-pulse fluctuation of the SC signal from water measured over 20 minutes exhibits an RMS $< 2 \cdot 10^{-3}$ for both pump wavelengths at the stability-optimized fluid velocity, which is comparable to the stability of YAG and sapphire, as shown in Fig. 3 a.2), b.2) and c.2). Since the cross-sectional area in the cell is 5.6 mm deep and 27.6 mm wide, the velocities of the fluid for pumping at 515 nm and 1030 nm are 3.7 mm/s and 2.1 mm/s respectively at the location where filamentation occurs. Pumping at longer wavelengths requires higher photon densities to generate a SC via multiphoton absorption (MPA) resulting in stronger Kerr focusing and thus a smaller filament radius in the nonlinear focus. A smaller plasma filament radius implies that the thermal energy is transferred to a smaller volume of surrounding water, in contrast to a larger radius at similar temperature during an equal period of time. This explains the reduction of the optimum fluid velocity for increasing pump wavelengths. Therefore, the volume of water that needs to be displaced depends on the characteristics of the formed plasma filament, and thus on the focusing conditions, the wavelength and the maximum peak power of the pulses.

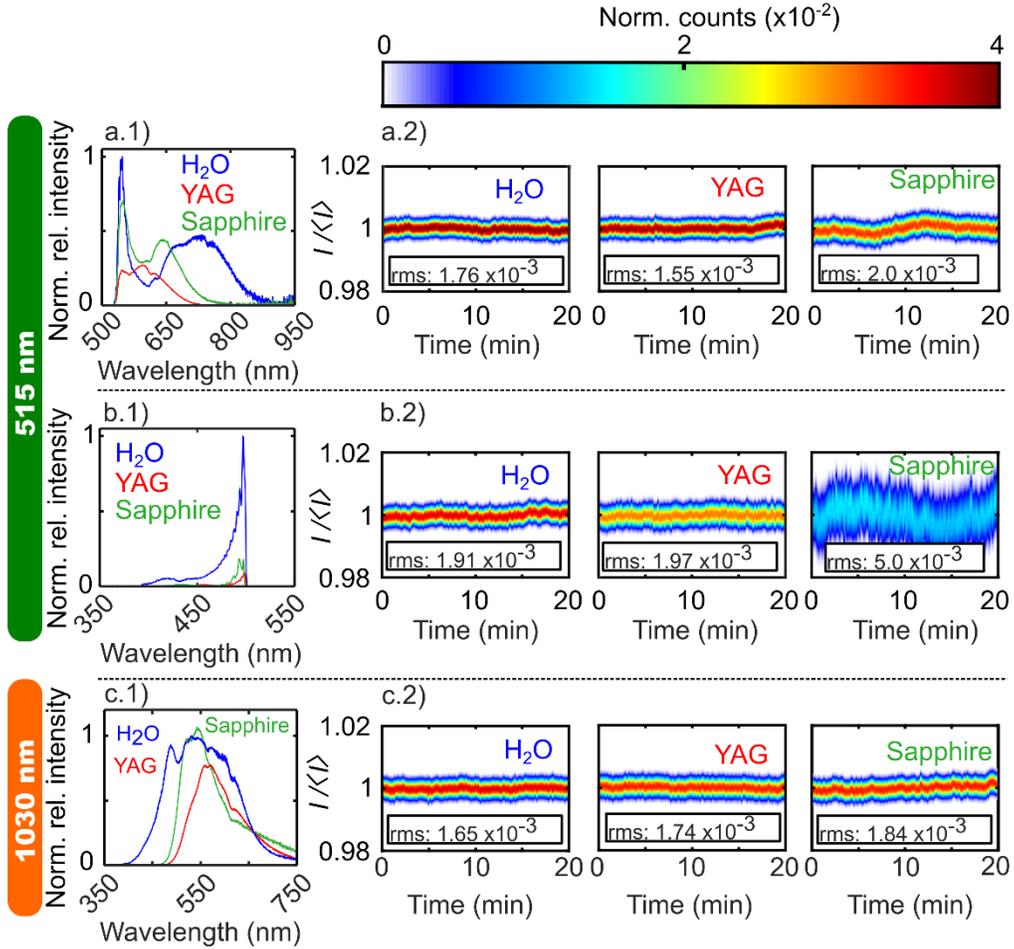

Fig. 3. Spectra a,b,c.1) and pulse-to-pulse fluctuations a,b,c.2) of the SC generated with the SH and the FW at 50 kHz in water (blue) with a thickness of 5.6 mm, and YAG (red) and sapphire (green) with a thickness of 6 mm each. The spectra and the stability with the SH pump are measured for the a.) Stokes part of the SC signal after a long pass filter and for the b.) anti-Stokes part after a short pass filter. For FW pump, the spectrum of the anti-Stokes part is plotted for the range below 750 nm while the stability is measured after a short pass filter. The SC spectra generated by the crystals are normalized to the maxima of the signals from water. The pulse-to-pulse fluctuations are visualized by stacked histograms, each representing the fluctuation of pulse intensity measured over one second as normalized relative counts.

Table 1. green: $\lambda_{5\%}$ wavelength and average power $P_{avg}$ at 700 nm and 450 nm of Stokes (ST) and anti-Stokes (AS) shoulder respectively for SH pump. orange: $\lambda_{5\%}$ wavelength and average power $P_{avg}$ at 671 nm of the anti-Stokes (AS) shoulder for FW pump.

| material | 515 nm | | | | 1030 nm | |
| --- | --- | --- | --- | --- | --- | --- |
| | $\lambda_{5\%,ST}$ | $\lambda_{5\%,AS}$ | $P_{avg,700\,nm}$ | $P_{avg,450\,nm}$ | $\lambda_{5\%,ST}$ | $P_{avg,671\,nm}$ |
| water | 860 nm | 415 nm | 10.5 µW | 41 µW | 415 nm | 6.9 µW |
| YAG | 716 nm | 442 nm | 0.7 µW | 7.4 µW | 492 nm | 7.7 µW |
| sapphire | 747 nm | 427 nm | 3.8 µW | 6.2 µW | 481 nm | 8.1 µW |

Fig. 4 a) shows the dependence of the root mean square (RMS) on the fluid velocity extracted from the pulse-to-pulse fluctuations of the SC generated in the flow cell at 50 kHz (circles) and at 100 kHz (squares) with a peak pump intensity of 690 GW cm$^{-2}$. Both repetition rates exhibit an RMS minimum below $0.15 \cdot 10^{-2}$, which is close to the RMS level of the laser pump source (dashed line in the figure), for velocities between 2 mm/s and 7 mm/s. In both cases the RMS value decreases drastically with increasing fluid velocity until it reaches the minimum value. The RMS then remains below $0.2 \cdot 10^{-2}$ up to a fluid velocity of 7 mm/s after which it increases linearly up to the maximum tested fluid velocity of 20 mm/s. The minimum RMS at 50 kHz occurs at a lower fluid velocity with respect to the corresponding minimum at 100 kHz. Based on the fact that the local water temperature increases faster at higher laser repetition rates, the observed behavior suggests that the instability for low fluid velocities rates has a thermal origin. A typical example of a thermally induced source of instability is optical cavitation and the resulting generation of bubbles, which can suppress the generation of white light [25]. On the other hand, to rule out a common source of instabilities at high and low fluid velocities, the noise measured in both cases with similar RMS is compared. Figure 4 a) shows for both repetition rates three measurements marked with different colours, at a low (yellow), optimum (red), and high (green) fluid velocity. Figure 4 b) and c) show the amplitude spectral density (ASD) for the cases marked in 4a) with repetition rates of 50 and 100 kHz, respectively. The following observations are valid for both repetition rates: the noise is strongly reduced for frequencies below 100 Hz in the cases with a constant fluid velocity compared to the case without flow. Among the cases with a constant flow, the optimum case provides similar or a lower noise level over the whole bandwidth. While the low and high fluid velocity cases have a comparable RMS, the former introduces less noise at frequencies around hundreds of Hertz compared to the latter but has a higher noise level at frequencies around tens of kHz. The difference in the spectral distribution of the noise between the two non-optimal flow cases suggests a different origin of the instabilities depending on the fluid velocity, namely thermal instabilities at low limit and the formation of turbulences in the liquid as a consequence of high fluid velocities. Therefore, precise control of the flow, and thus of the fluid velocity, as achieved with the proposed scheme, is crucial in order to optimize the stability of the white light signal generated in liquids with optical pulses at high repetition rates.

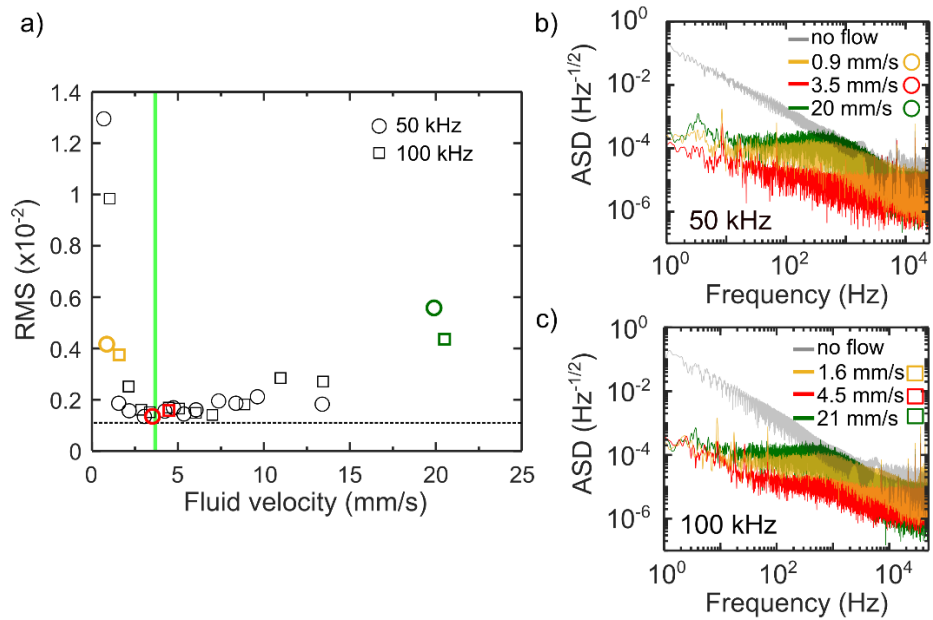

Fig. 4. RMS of the pulse-to-pulse fluctuations of the SC generated by the SH in water as a function of fluid velocity inside the cell for a repetition rate of 50 kHz (red) and 100 kHz (blue). The green bar indicates the fluid velocity of 3.7 mm/s used for the measurements at 50 kHz in Fig. 3. The black dotted line represents the RMS of the pump laser. The colored datapoints indicate the RMS values obtained for low (yellow), optimum (red) and high (green) fluid velocity respectively. b) and c) show the amplitude spectral density (ASD) of the pulse train for 50 kHz and 100 kHz repetition rates respectively when there is no flow (grey) and with the low (yellow), optimum (red), and high (green) fluid velocity of the measurements indicated in a).

The ideal nonlinear medium for SC generation combines high efficiency for spectral broadening, pump-source-limited stability, high damage threshold, and flexibility with respect to the pump wavelength and the repetition rate of the driving pulses. In this work we developed an experimental setup which allows to harness water as an effective medium for octave spanning SC generation and significantly exceed the performance of well-established nonlinear materials such as YAG and sapphire. The setup presented here allows to establish a nearly constant flow of water inside a liquid cell and to fine tune the velocity by means of a differential pressure control in order to overcome the main challenges for stable SC generation in liquids due to the thermal phenomena induced by intense pulses at multi-kHz repetition rates. With a careful analysis of how instabilities at multi-kHz repetition rates depend on the laser parameters, their dependence on the fluid velocity and the spectral noise distribution, we prove a viable platform for the exploitation of water as an effective nonlinear medium for operations even at repetition rates as high as 100 kHz. This approach is quite flexible since it can be adapted to different pump wavelengths and since it is intrinsically robust when compared to alternative approaches such as liquid jets, that are much more complicated to optimize and do not provide sufficient interaction lengths. Remarkably, the technical solutions introduced here can be easily implemented with other liquid media that offer a wider range of transparency than water, such as bromine [26] or pure carbon disulfide (CS2) [26], with the perspective of expanding the availability of stable liquid-based SC up to the mid-infrared.

**Funding.** European Regional Development Fund (2017-03-022-19); European Research Council (819871)

**Disclosures.** The authors declare no conflicts of interest.

**Data availability.** Data underlying the results presented in this paper are not publicly available at this time but may be obtained from the authors upon reasonable request.